\documentclass[3p,times,procedia]{elsarticle}
\usepackage{nupha_ecrc}
\def\bea#1\eea{\begin{align}#1\end{align}}

\newcommand{\bef}{\begin{figure}[!htp]}
\newcommand{\eef}{\end{figure}}

\volume{00}

\firstpage{1}

\journalname{Nuclear Physics A}

\runauth{}

\jid{nupha}

\jnltitlelogo{Nuclear Physics A}

\usepackage{amssymb}

\usepackage[figuresright]{rotating}

\begin{document}

\begin{frontmatter}



\dochead{XXVIIIth International Conference on Ultrarelativistic Nucleus-Nucleus Collisions\\ (Quark Matter 2019)}

\title{Extracting the jet transport coefficient of cold nuclear matter from world data}


 \author[label1]{Peng Ru}
 \author[label2,label3]{Zhong-Bo Kang}
 \author[label1]{Enke Wang}
 \author[label1]{Hongxi Xing$^{*,}$\corref{cor}}
 \author[label4]{Ben-Wei Zhang}

 \cortext[cor]{Speaker}

 \address[label1]{Guangdong Provincial Key Laboratory of Nuclear Science, Institute of Quantum Matter,
 South China Normal University, Guangzhou 510006, China}
 \address[label2]{Department of Physics and Astronomy, University of California, Los Angeles, California 90095, USA}
 \address[label3]{Mani L. Bhaumik Institute for Theoretical Physics, University of California, Los Angeles, California 90095, USA}
 \address[label4]{Key Laboratory of Quark $\&$ Lepton Physics (MOE) and Institute of Particle Physics, Central China Normal University, Wuhan 430079, China}

\address{}

\begin{abstract}
We present the first global extraction of the jet transport coefficient~($\hat q$) for cold nuclear matter within the
framework of higher-twist expansion.
The analysis takes into account the world data on transverse momentum broadening in semi-inclusive $e$-A deep inelastic scattering (DIS) and in Drell-Yan dilepton and heavy quarkonium production in $p$-A collisions, as well as the nuclear modification of the structure functions in DIS. The results of this work provide the first quantitative evidence for the universality and non-trivial scale dependence of the medium transport property, similar to that for nucleon structure as encoded in the standard parton distribution functions. We expect the extracted scale dependence of $\hat q$ for cold nuclear matter can be extended to precisely identify the fundamental property of quark gluon plasma.
\end{abstract}


\end{frontmatter}


\section{Introduction}
\label{Introduction}
Quantifying the nuclear modification due to multiple scatterings between the hard probe and nuclear medium
can provide a solid baseline for the identification of the medium fundamental property.
In relativistic nucleus-nucleus~(AA) collisions, the interaction between the hard probe and nuclear medium leads to nontrivial phenomenon, such as jet quenching which sheds light on the property of
the hot/dense nuclear medium~\cite{Burke:2013yra}.
On the other hand, electron-nucleus~($e$A) and proton-nucleus~($p$A) collisions provide a clean environment
to study the jet transport property of cold nuclear medium. A reliable framework that has been extensively used in both hot/dense medium and cold nuclear matter is the so-called higher-twist factorization formalism, in which the medium property is encoded in the non-perturbative twist-4 matrix element, which in turn can be converted into the jet transport coefficient $\hat q$. 
$\hat q$ represents the transverse momentum broadening per unit propagation length of a jet encountered in the medium, and characterizes the strength of the jet-medium interaction.

As a crucial input of the jet-quenching study, $\hat q$ has been widely used as a constant, i.e., a probing-scale
independent quantity. However, the recent theoretical developments of the renormalization group equations~\cite{Kang:2013raa,Kang:2014ela,Kang:2016ron,Blaizot:2014bha,Iancu:2014kga,Liou:2013qya}
for $\hat q$ and the next-to-leading~(NLO) order calculations of the transverse momentum broadening for the real scattering
processes in cold nuclear medium\cite{Kang:2013raa,Kang:2014ela,Kang:2016ron} have indicated the universality and scale dependence of $\hat q$.
Although higher-order calculations are needed to make the evolution equation for $\hat q$ to be closed,
at this stage, a data-driven study of the $\hat q$ in cold nuclear medium will be meaningful for understanding
the probing-scale dependence of the nuclear medium property.

\section{Phenomenology for $\hat q$ within framework of higher-twist expansion}
In this work~\cite{Ru:2019qvz}, we perform the first global extraction of the $\hat q$ for cold nuclear matter within the
theoretical framework of the generalized QCD factorization formalism, i.e., the higher-twist expansion,
which has been shown to be a successful approach to describe the nuclear effects observed in heavy ion collisions.

For the study of $\hat q$ for cold nuclear matter, of particular relevance is the phenomenon of transverse
momentum~($p_T$) broadening. Take semi-inclusive $e$A deep inelastic scattering~(SIDIS) as an example,
the $p_T$ broadening of the final-state hadron defined as
$\Delta \langle p_T^2\rangle \equiv \langle p_T^2\rangle _{eA} - \langle p_T^2\rangle_{ep}$
can be approximated by the leading contribution from final state double scattering~\cite{Guo:1998rd}
\begin{equation}
\Delta \langle p_T^2\rangle = \frac{4\pi^2\alpha_sz_h^2}{N_c}
\frac{\sum_q T_{qg}(x_B,0,0,\mu^2) D_{h/q}(z_h,\mu^2)}{\sum_q f_{q/A}(x_B,\mu^2) D_{h/q}(z_h,\mu^2)}\,,
\label{eq-dis}
\end{equation}
where $f_{q/A}(x_B,\mu^2)$ and $D_{h/q}(z_h,\mu^2)$ represent the usual leading twist parton distribution function and the hadron fragmentation function,
and $T_{qg}(x_B,0,0,\mu^2)$ is the twist-4 quark-gluon correlation function which can be effectively factorized as \cite{CasalderreySolana:2007sw}
\begin{equation}
T_{qg}(x,0,0,\mu^2) \approx \frac{9R_A}{8\pi^2\alpha_s} f_{q/A}(x,\mu^2)\, \hat q(x,\mu),
\label{eq:Tqg}
\end{equation}
with $R_A$ the nuclear radius and $\hat q(x,\mu)$ the averaged quark jet transport coefficient in cold nuclear matter.
In this global analysis, we take into account the experimental measurements of the $p_T$ broadening in SIDIS and
in Drell-Yan dilepton and heavy-quarkonium production in $p$A collisions, which involve the quark- and gluon-initiated
double scattering processes in initial and final states of the collisions~\cite{Kang:2013raa,Kang:2008us}.
Besides, part of the measurement of the nuclear modification of the DIS structure functions related to dynamical shadowing
is also included in the analysis, which can be calculated by resuming the high-twist contributions~\cite{Qiu:2003vd}.

\section{Global analysis and results}
\begin{table}[h]
\hspace{1cm}
\begin{tabular}{lcccc}
experiment & data type & data points & $\chi^2$~~(constant $\hat q$) & $\chi^2$~~[$\hat q\,(x_B,Q^2)$] \\ \hline
HERMES   & SIDIS~~($p_T$ broad.)     &    156  & 218.5  & 189.7\\
FNAL-E772    & DY~~($p_T$ broad.)     &    4  & 2.69  &  1.65 \\
SPS-NA10     & DY~~($p_T$ broad.)     &   5  &  6.86 & 6.47 \\
FNAL-E772  & $\Upsilon$~~($p_T$ broad.)     &   4  & 2.33  & 2.67 \\
FNAL-E866  & $J/\psi$~~($p_T$ broad.)    &   4  &  2.03 & 2.45 \\
RHIC   & $J/\psi$~~($p_T$ broad.)     &    10  & 44.4 & 31.0 \\
LHC    & $J/\psi$~~($p_T$ broad.)    &    12  & 87.3 & 4.8 \\
FNAL-E665    & DIS~~(shadowing)  &  20  & 23.7 & 21.46
    \\ \hline
{\bf TOTAL:} & & 215 & 387.9 & 260.2\\
\end{tabular}
\caption{\label{table1}
Data sets used in the global analysis, and the $\chi^2$ values with a constant $\hat q$ and $\hat q(x_B,Q^2)$, respectively.
}
\end{table}

\begin{figure}[t]
\label{krange}
\includegraphics[scale=0.45]{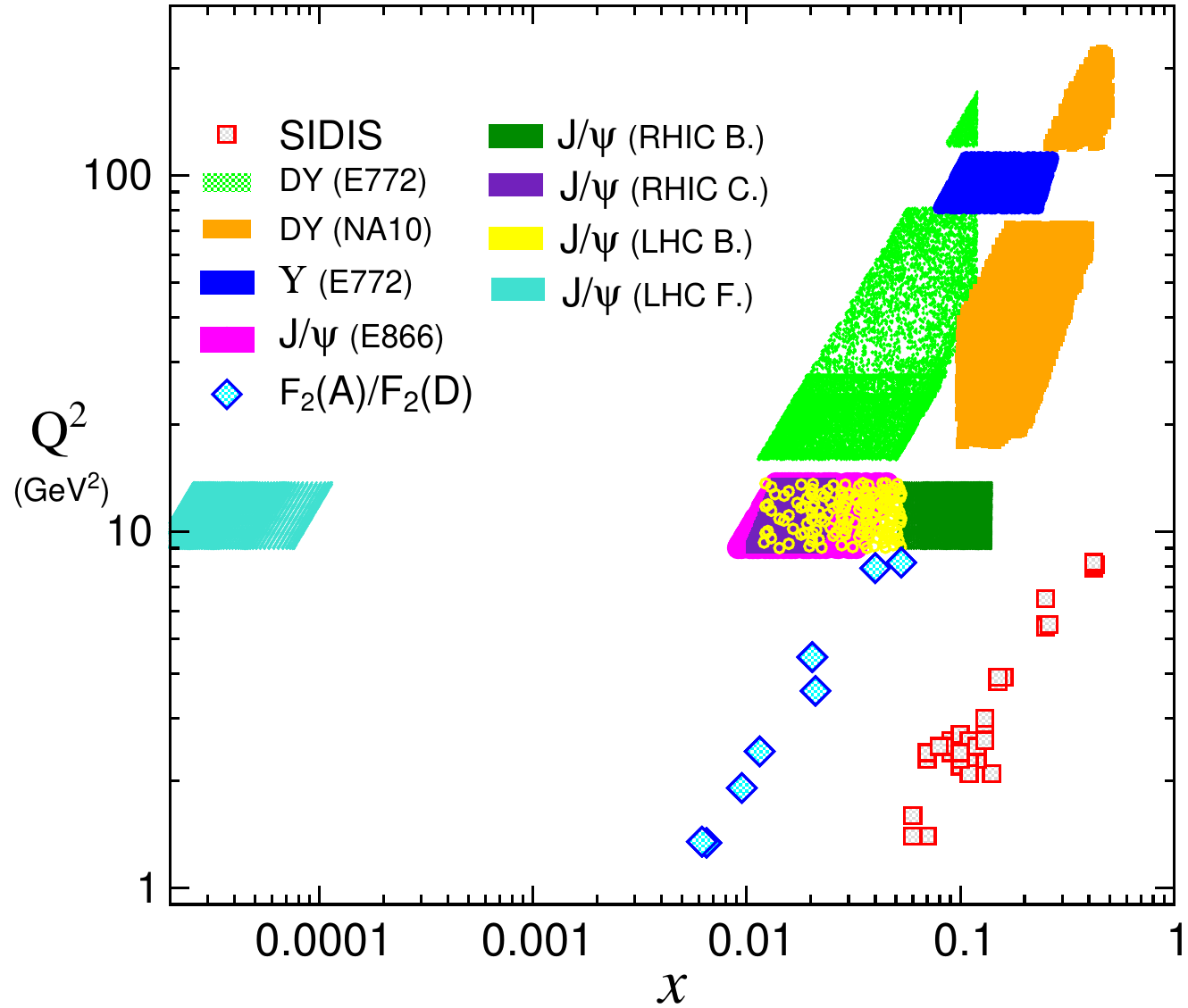}
\hspace{1.6cm}
\includegraphics[scale=0.4]{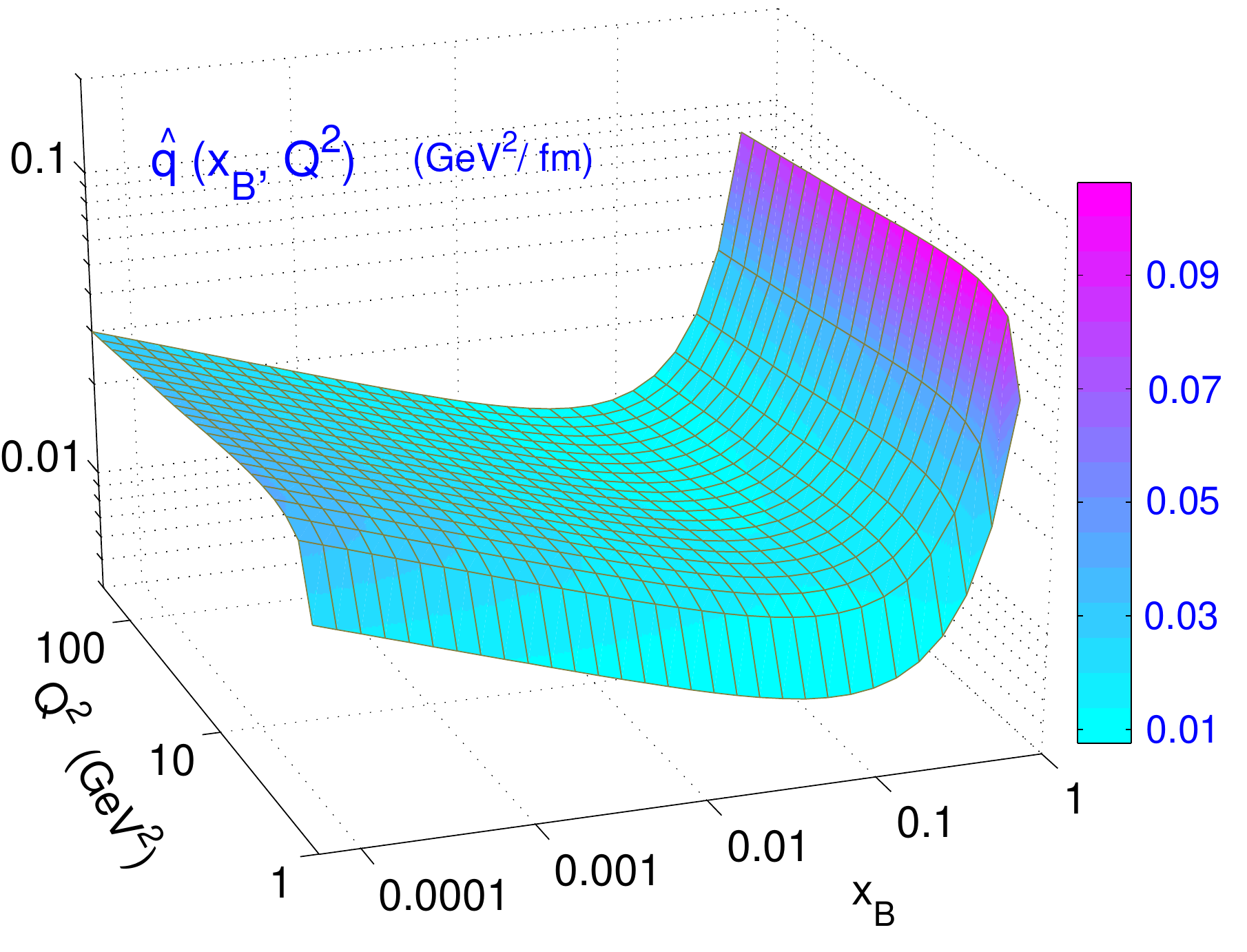}
\caption{Range of Bjorken-$x$ and $Q^2$ accessed by the used data~(left panel) and extracted central values of $\hat q\,(x_B,Q^2)$~(right panel).}
\label{fig:xq}
\end{figure}

The data sets used in this global analysis are listed in Table.~\ref{table1},
while the range of the Bjorken $x$~($x_B$) and $Q^2$ accessed by the used data is shown in the left panel of Fig.~\ref{fig:xq}.
Through the analysis for each individual data set with $\hat q$ assumed to be a constant quantity,
we find that the extracted $\hat q$ values can even differ by a factor of 2-4~\cite{Ru:2019qvz},
indicating the necessity of a nontrivial probing scale dependence of $\hat q$ for describing various data simultaneously.
To properly consider the $x_B$ and $Q^2$ dependence of $\hat q$, we utilize the parametrization form as
\begin{equation}
\hat q(x_B,Q^2) = \hat q_0 \,\alpha_s(Q^2) \,x_B^{\alpha}(1-x_B)^{\beta} \left[\ln(Q^2/Q_0^2)\right]^{\gamma}\,.
\label{eq-pramt}
\end{equation}
In Table.~\ref{table1}, we compare the $\chi^2$ values obtained in the global fitting with a constant $\hat q$ 
to those with the $\hat q\,(x_B,Q^2)$ in Eq.~(\ref{eq-pramt}).
We find that the total $\chi^2$ is significantly reduced by applying the parametrization form in Eq.~(\ref{eq-pramt}).
The extracted parameters for $\hat q\,(x_B,Q^2)$ are $\hat{q}_0=0.0195^{+0.0020}_{-0.0020}$~GeV$^2$/fm, $\alpha=-0.174^{+0.031}_{-0.025}$,
$\beta=-2.79^{+0.91}_{-0.80}$, and $\gamma=0.254^{+0.132}_{-0.143}$, where the uncertainties correspond to $90\%$ confidence level.
The extracted central values of $\hat q\,(x_B,Q^2)$ are shown in the right panel of Fig.~\ref{fig:xq}.
These results provide a quantitative evidence for a nontrivial kinematic dependence of the $\hat q$ in cold nuclear matter.

\begin{figure}[t]
\includegraphics[scale=0.62]{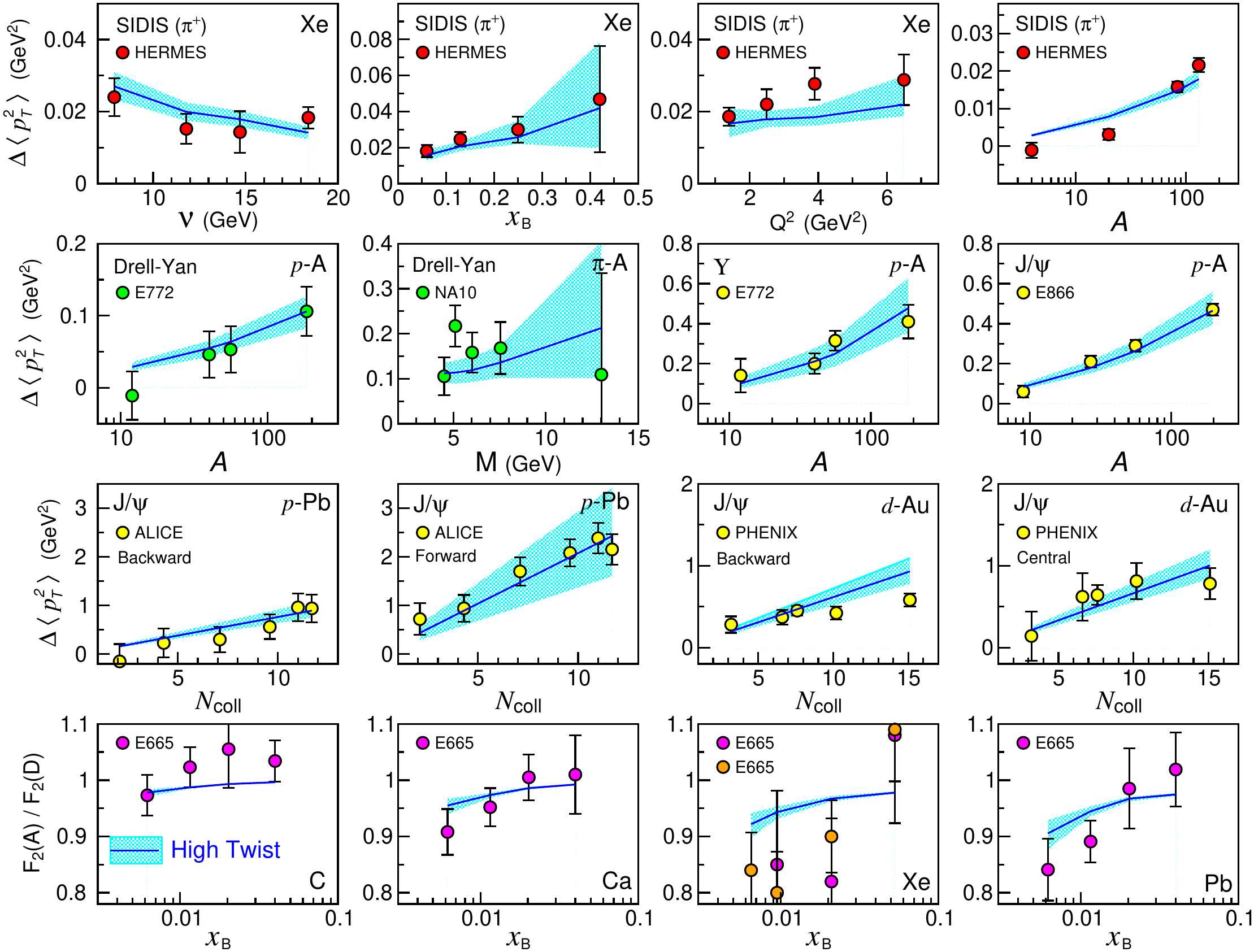}
\caption{Global fit compared to experimental data. Shown are representative comparisons.}
\label{fig:fit}
\end{figure}

In Fig.~\ref{fig:fit} we show the comparison between the theoretical results corresponding to the
extracted $\hat q\,(x_B,Q^2)$ and the representative data for the $p_T$ broadenings in SIDIS, Drell-Yan
process and heavy quarkonium production, as well as for the shadowing in nuclear structure functions.
Clearly, the higher-twist calculations can well describe various types of data simultaneously.
For SIDIS, the calculations capture the $p_T$ broadening increasing with $x_B$ in the region $x_B\gtrsim0.1$,
owing to the increasing $\hat q$ as shown in the right panel of Fig. \ref{fig:xq}.
Comparing the $J/\psi$ production at the LHC in the forward region to that in the backward region, we find that the calculations
can well describe the $p_T$ broadening decreasing with increasing Bjorken $x$ at smaller $x$ region, which is related to the behavior of $\hat q$ in small $x$ region shown in Fig. \ref{fig:xq}.
The typical linear medium size dependence in the higher-twist formalisms~[see Eq.~(\ref{eq:Tqg})] agrees well with the data.
Although this global analysis is dominated by the various $p_T$ broadening data, it is interesting to observe that
the higher-twist results are in agreement with the data for dynamic shadowing effect.
Besides, the uncertainties of the theoretical predictions from the extracted $\hat q\,(x_B,Q^2)$ vary with the observables,
indicating that the constraining powers of the experimental data vary with the underlying $x_B$ and $Q^2$ regions.

\section{Summary}
In summary, we have performed the first global extraction of the jet transport coefficient for
cold nuclear matter within the framework of higher-twist expansion. Our results provide a quantitative
evidence that the $\hat q$ for cold nuclear matter is an apparently probing-scale dependent quantity,
rather than a constant value usually used in heavy ion studies, which is expected to motivate a more
precise understanding of the jet transport property of the quark-gluon plasma. The LO results of this
work may be improved in future when NLO theoretical framework and numerical implementation are available. 
On the other hand, the data sets used in this analysis is primarily the $p_T$ broadening with the 
$x_B$ and $Q^2$ located in intermediate region, future measurement with more types of observable and a 
wider kinematic range, e.g., at the EIC, will provide valuable constraints on the $\hat q$ for cold nuclear matter.

\section{Acknowlegement}
This work is supported in part by China Postdoctoral Science Foundation under project No. 2019M652929~(P.R.),
by the National Science Foundation in US under Grant No. PHY-1720486~(Z.K.), and by NSFC of China under Project
No. 11435004~(E.W., H.X. and B. Z.).





\bibliographystyle{elsarticle-num}
\bibliography{<your-bib-database>}



\end{document}